\documentclass[%
reprint,
superscriptaddress,
groupedaddress,
unsortedaddress,
runinaddress,
frontmatterverbose, 
showpacs,
preprintnumbers,
nofootinbib,
nobibnotes,
amsmath,amssymb,
aps,
pra,
prb,
rmp,
prstab,
prstper,
floatfix,
]{revtex4-1}
\usepackage{graphicx}
\usepackage{dcolumn}
\usepackage{bm}
\usepackage[mathlines]{lineno}
\begin{document}
\preprint{}
\title{Classical analytic solutions of the non-stationary Navier-Stokes equation in two, three and higher dimensions}
\author{R. K. Michael Thambynayagam}\altaffiliation{Schlumberger, Houston, Texas, USA}
 \email{thamby@slb.com}


\date{\today}
\begin{abstract}
In this paper we present a method to derive classical solutions of the Navier-Stokes equations for non-stationary initial  value problems  in domain $\mathbb{R}^n$ ($n=2,3$ or higher). Exact solutions in $\mathbb{R}^2$ and $\mathbb{R}^3$ in the presence of an externally specified force are obtained by imposing certain restrictions on the initial data.
%
%
\end{abstract}

\maketitle


\section{The fundamental problem}
The motion of incompressible Newtonian fluids are described by the Navier-Stokes equations\cite{Nav, Sto}. These equations are an expression of Newton's second law, and in the presence of an external force they are
%
\begin{eqnarray}
\frac{{\partial v_i }}
{{\partial t}} + g_i= \kappa \Delta v_i -\frac{1}
{\rho }\frac{{\partial p}}
{{\partial x_i }} + f_i\quad {x \in \mathbb{R}^n},\quad t \geqslant 0\quad 
\end{eqnarray}
\vspace{-0.20in}
%
\begin{eqnarray}
\protect{\bf{div}}\,v = \sum\limits_{i = 1}^n {\frac{{\partial v_i }}
{{\partial x_i }}}=0 \qquad
x \in \mathbb{R}^n,\,\,t \geqslant 0
\end{eqnarray}
where 
\vspace{-0.20in}
\begin{eqnarray}
g_i  = \sum_{j = 1}^n {v_j \frac{{\partial v_i }}
{{\partial x_j }}} 
\end{eqnarray}
is the nonlinear inertial force, $v_i\left( {x,t} \right)$ is the velocity field evaluated at point $x \in \mathbb{R}^n$ at time $t \geqslant 0$, $p$ is the pressure field, $f_i\left( {x,t} \right)$ are the components of an externally applied force, $\rho$ is the constant density of the fluid,  $\kappa$ is the positive coefficient of kinematical viscosity and $\Delta  = \sum_{i = 1}^n {\frac{{\partial ^2 }}{{\partial x_i^2 }}}$ is the Laplacian in the space variables.\\\\ The initial conditions are
%
\begin{eqnarray}
v_i\left( {x,0} \right) = v_i^0 
 \left( x \right)\qquad
x \in \mathbb{R}^n 
\end{eqnarray}
$v_i^0 \left( x \right)$ is a given, $C^\infty $ divergence-free vector field on $\mathbb{R}^n$. Henceforth, the superscript $0$ is used to denote the value of a function at time zero. It is important to note that prescribing pressure at the initial time independent of velocity would render the problem ill-posed.\\\\
Conservation law implies that the energy dissipation of a viscous fluid is bounded by the initial kinetic energy which is finite and therefore the solution must satisfy
%
\begin{eqnarray}
\int\limits_{\mathbb{R}^n } {\left| {v\left( {x,t} \right)} \right|} ^2 dx < \mathcal{C}\qquad t \geqslant 0
\end{eqnarray}
The objective of the present paper is to determine exact  solutions of $v\left( {x,t} \right)$, $p\left( {x,t} \right)$ to the system of equations $\left(1\right)-\left(4\right)$.
\section{A solution of the problem\\ $\qquad\mathbb{R}^n=\left\{ { - \infty  < x_i  < \infty ; \,\, i = 1,2,...,n} \right\}$}
Assuming that the divergence and the linear operator can be commuted, the pressure field can be formally obtained by taking the divergence of equation $\left(1\right)$ as a solution of the Poisson equation, which is
%
\begin{eqnarray}
\Delta p = \rho \sum\limits_{i = 1}^n {\frac{{\partial \left( {f_i  - g_i } \right)}}
{{\partial x_i }}} 
\end{eqnarray}
Equation $\left(6\right)$ is called the simplified pressure Poisson equation (PPE). The use of PPE in solving the Navier-Stokes equation is discussed in an illuminating paper by Gresho \& Sani\cite{gre}. It is important to note that  while equations $\left(1\right)$ and $\left(2\right)$ lead to the pressure Poisson equation $\left(6\right)$, the reverse; that is, equations $\left(1\right)$ and $\left(6\right)$, do not always lead to equation $\left(2\right)$. We therefore, in deriving  exact solutions of the Navier-Stokes equations, ensure that  the velocity vector field remains solenoidal at all times. \\\\ The general solution of the Poisson equation $\left(6\right)$ is
%
%
\begin{eqnarray}
p &=&- \frac{\rho }
{{2\pi }}\int\limits_{\mathbb{R}^2 } {{\rm P}\left( {y ,t} \right)\ln \left( {\frac{1}
{{\sqrt {\mathcal{P}_n\left( {x,y} \right)} }}} \right)\!\prod\limits_{j = 1}^2 {dy_j } },\,\, n=2, \nonumber\\
p &=&- \frac{{\rho \Gamma \left( {\frac{n}
{2}} \right)}}
{{2\left( {n - 2} \right)\pi ^{\frac{n}
{2}} }}\int\limits_{\mathbb{R}^n }\! {\frac{{{\rm P}\left( {y,t} \right)}}
{{\left\{ {\mathcal{P}_n\left( {x,y} \right)} \right\}^{\frac{{n - 2}}
{2}} }}} \!\prod\limits_{j = 1}^n \!{dy_j },\,\, n\geqslant 3\,\,\, 
\end{eqnarray}
where $\Gamma \left( z \right) = \int_0^\infty  {e^{ - u} u^{z - 1} du}$\quad$\left[ {\Re z > 0} \right]$, is the Gamma function,
%
\begin{eqnarray}
{\rm P}\left( {x,t} \right) = \sum\limits_{j = 1}^n {\frac{{\partial \left( {f_j  - g_j } \right)}}
{{\partial x_j }}} 
\end{eqnarray}
and
%
%
\begin{eqnarray}
\mathcal{P}_n\left( {x,y} \right) = \sum\limits_{j = 1}^n {\left( {x_j  - y_j } \right)^2 } 
\end{eqnarray}
Differentiating equation $\left(7\right)$ with respect to $x_i$ we get 
%
%
\begin{eqnarray}
\frac{{\partial p }}
{{\partial x_i }} = \frac{{\rho \Gamma \left( {\frac{n}
{2}} \right)}}
{{2\pi ^{\frac{n}
{2}} }}\int\limits_{\mathbb{R}^n } {\frac{{\left( {x_i  - y_i } \right){\rm P}\left( {y,t} \right)}}
{{\left\{ {\mathcal{P}_n\left( {x,y } \right)} \right\}^{\frac{n}
{2}} }}} \prod\limits_{j = 1}^n {dy_j },\,\, n\geqslant 2\qquad  
\end{eqnarray}
Substituting for $\frac{{\partial p}}{{\partial x_i }}$ in equation $\left(1\right)$ we get
%
\begin{eqnarray}
\frac{{\partial v_i }}
{{\partial t}} &=&\kappa \Delta v_i  -\frac{{\Gamma \left( {\frac{n}
{2}} \right)}}
{{2\pi ^{\frac{n}
{2}} }}\int\limits_{\mathbb{R}^n } {\frac{{\left( {x_i  - y_i } \right){\rm P}\left( {y,t} \right)}}
{{\left\{ {\mathcal{P}_n\left( {x,y} \right)} \right\}^{\frac{n}
{2}} }}} \prod\limits_{j = 1}^n {dy_j }+\nonumber\\ 
&+&f_i - g_i\quad n\geqslant 2,\quad {x \in \mathbb{R}^n},\quad t \geqslant 0\quad  
\end{eqnarray}
The difficulty in solving the system of equations $\left(1\right)-\left(4\right)$ stems from the presence of the nonlinear term $g_i$.  We therefore recast the Navier-Stokes equation $\left(11\right)$ as:
%
\begin{eqnarray}
\frac{{\partial v_i }}
{{\partial t}} &=&\kappa \Delta v_i +\mathcal{F}_i\!\left( {x ,t} \right)  - \mathcal{U}_i\left( v \right)\quad {x \in \mathbb{R}^n},\quad t \geqslant 0\quad
\end{eqnarray}
\vspace{-0.20in}
where
%
\begin{eqnarray}
\mathcal{U}_i\left( v \right)  = g_i  - \frac{{\Gamma \left( {\frac{n}
{2}} \right)}}
{{2\pi ^{\frac{n}
{2}} }}\!\int\limits_{\mathbb{R}^n }\! {\frac{{\left( {x_i  - y_i } \right)\!\sum\limits_{k= 1}^n {\frac{{\partial g_k \left( {y,t} \right)}}
{{\partial y_k }}} }}
{{\left\{ {\mathcal{P}_n\left( {x,y} \right)} \right\}^{\frac{n}
{2}} }}}\! \prod\limits_{j = 1}^n\! {dy_j },\nonumber\\\quad {x \in \mathbb{R}^n},\,\, t \geqslant 0 \qquad
\end{eqnarray}
\begin{eqnarray}
\mathcal{F}_i\!\left( {x ,t} \right)  = f_i  - \frac{{\Gamma \left( {\frac{n}
{2}} \right)}}
{{2\pi ^{\frac{n}
{2}} }}\!\!\int\limits_{\mathbb{R}^n }\! {\frac{{\left( {x_i  - w_i } \right)\!\sum\limits_{k = 1}^n \!{\frac{{\partial f_i \left( {w ,t} \right)}}
{{\partial w_k }}} }}
{{\left\{ {\mathcal{P}_n\left( {x,w} \right)} \right\}^{\frac{n}
{2}} }}}\! \prod\limits_{j = 1}^n\! {dw_j },\nonumber\\\quad {x \in \mathbb{R}^n},\,\, t \geqslant 0 \qquad
\end{eqnarray}
$v_i  \equiv v_i \left( {x_1 ,x_2 ,...,x_n ,t} \right)$, $\mathcal{U}_i \left( v \right)\equiv\mathcal{U}_i \left( {v_1 ,v_2 ,...v_n } \right)$ and $\mathcal{F}_i \left( {x,t} \right) \equiv \mathcal{F}_i \left( {x_1 ,x_2 ,...,x_n ,t} \right)$. Due to the solenoidal nature of the velocity field, the initial form of $\mathcal{U}_i \left( v \right)$ is preserved at all times.\\\\
The three terms on the right hand side of equation $\left(12\right)$ are associated, respectively, with the linear viscous force, the externally applied force and the nonlinear inertial force. The nonlinearity of the problem is isolated within $\mathcal{U}_i\left( v \right)$ which we, henceforth, refer to as the \protect{\itshape{umbilical force}}. The formulation separates the smooth and non-smooth parts of the Navier-Stokes equation. The latter, the nonlinear part, may render the solution to have discontinuities\cite{num, kim, far}.
\subsection{\protect{\large{Exact solutions}}}
A posteriori we state that there exists an exact solution of the Navier-Stokes equation in $\mathbb{R}^n$ if the umbilical force $\mathcal{U}_i \left( v \right)$, defined by equation $\left(13\right)$, is either a function of time  only or equal to zero.\\\\ If the  initial conditions $v_i^0  \left( x \right)$ and the components of the externally applied force $f_i\left( {x,t} \right)$ are chosen such that the umbilical force 
$\mathcal{U}_i \left( v \right)\equiv {\mathcal{A}_i}\left( t \right)$, a function of time only, and the incompressibility of the fluid is enforced at all times, then, the solution of the non-homogeneous diffusion equation\cite{tha2} 
%
\begin{eqnarray}
v_i &=&\frac{1}
{{\left( {2\sqrt {\pi \kappa t} } \right)^n }}\!\int\limits_{\mathbb{R}^n } {v_i^0 \left( {y} \right)e^{ - \sum\limits_{k = 1}^n {\frac{{\left( {x_k - y_k } \right)^2 }}
{{4\kappa t}}} } \prod\limits_{j = 1}^n {dy_j } }  + \nonumber\\
&+&\frac{1}{\left( {2\sqrt {\pi \kappa } } \right)^n }\!\!\int\limits_{\mathbb{R}^n }\! {\int\limits_0^t {\frac{{\mathcal{F}_i \left( {y ,\tau } \right)e^{ - \sum\limits_{k = 1}^n {\frac{{\left( {x_k  - y_k } \right)^2 }}
{{4\kappa \left( {t - \tau } \right)}}} } }}
{{\left( {t - \tau } \right)^{\frac{n}{2}}}}d\tau\prod\limits_{j = 1}^n {dy_j } } }\nonumber\\
&+&\int\limits_0^t {{\mathcal{A}}_i \left( \tau  \right)} d\tau 
\end{eqnarray}
is a solution of the Navier-Stokes equation $\left(12\right)$.\\\\ 
We make the following remarks on equation $\left(15\right)$:\\\\
$\left(\rm{i}\right)$ In the $\mathop {\lim }\limits_{t \to 0} v_i$ the first term on the right hand-side of equation $\left(15\right)$ approaches $v_i^0$ and the second and the third terms vanishes.\\\\
$\left(\rm{ii}\right)$ The functional form of the recurring term ${e^{ - \sum_{k = 1}^n {\frac{{\left( {x_k  - y_k } \right)^2 }}{{4\kappa t}}} } }$  is $C^\infty$ in $\left( {x ,t} \right)$, $t\geqslant0$, which implies real analytic control on $v_i$. The solution is smooth.\\\\ 
$\left(\rm{iii}\right)$ In order to satisfy $\left(5\right)$, $\int_0^t {\mathcal{A}\left( \tau  \right)} d\tau$ must be integrable and bounded in any given interval of time.\\\\
$\left(\rm{iv}\right)$ If $f_i\left( {x,t} \right)={\mathcal{G}_i}\left( t \right)$, a function of time only, then, the second term on the right hand side of equation $\left(15\right)$ may be replaced by $\int_0^t {{\mathcal{G}}_i \left( \tau  \right)} d\tau$. In order to satisfy $\left(5\right)$, $\int_0^t {{\mathcal{G}}_i \left( \tau  \right)} d\tau$ must be integrable and bounded in any given interval of time.\\\\ 
$\left(\rm{v}\right)$ Since the \protect{\bfseries{div}} $v_i^0=0$, $f_i \left( {x,t} \right) = {\mathcal{G}_i}\left( t \right)v_i^0 \left( x \right)$ is also divergence free.\\\\
$\left(\rm{vi}\right)$ If $\mathcal{F}_i \left( {x,t} \right)$ set to zero, that is
%
\begin{eqnarray}
 f_i = \frac{{\Gamma \left( {\frac{n}
{2}} \right)}}
{{2\pi ^{\frac{n}
{2}} }}\!\!\int\limits_{\mathbb{R}^n }\! {\frac{{\left( {x_i  - w_i } \right)\!\sum\limits_{k = 1}^n \!{\frac{{\partial f_i \left( {w ,t} \right)}}
{{\partial w_k }}} }}
{{\left\{ {\mathcal{P}_n\left( {w,x} \right)} \right\}^{\frac{n}
{2}} }}}\! \prod\limits_{j = 1}^n\! {dw_j }\qquad 
\end{eqnarray}
then,  the second term on the right hand-side of equation (15) vanishes and the solution of the Navier-Stokes equation (12) reduces to that of the Cauchy diffusion equation\cite{ch1}, which is
%
\begin{eqnarray}
v_i &=&\frac{1}
{{\left( {2\sqrt {\pi \kappa t} } \right)^n }}\!\int\limits_{\mathbb{R}^n } {v_i^0 \left( {y} \right)e^{ - \sum\limits_{k = 1}^n {\frac{{\left( {x_k - y_k } \right)^2 }}
{{4\kappa t}}} } \prod\limits_{j = 1}^n {dy_j } } 
\end{eqnarray}
We verify the authenticity of the aforementioned assertions by deriving exact solutions in $\mathbb{R}^2$ and $\mathbb{R}^3$, though in $\mathbb{R}^2$, similar assertions have been known for a long time\cite{lad}.\\\\ For the sake of simplicity we have considered the umbilical force $\mathcal{U}_i \left( v \right) \equiv 0$; that is,
%
\begin{eqnarray}
g_i = \frac{{\Gamma \left( {\frac{n}
{2}} \right)}}
{{2\pi ^{\frac{n}
{2}} }}\!\int\limits_{\mathbb{R}^n }\! {\frac{{\left( {x_i  - y_i } \right)\!\sum\limits_{k = 1}^n {\frac{{\partial g_k }}
{{\partial y_k }}} }}
{{\left\{ {\mathcal{P}_n\left( {x,y} \right)} \right\}^{\frac{n}
{2}} }}}\! \prod\limits_{j = 1}^n\! {dy_j }
\end{eqnarray}
\begin{center}
\protect{{\bfseries{Exact solutions in $\mathbb{R}^2$\\ $\mathbb{R}^2=\left\{ { - \infty  < x_i  < \infty ; \,\, i = 1,2} \right\}$}}}
\end{center}
\protect{\small{\bfseries{Problem$\left(\rm{i}\right)$}}} $v_1^0  = \sin \left( {\pi x_1 } \right)\cos \left( {\pi x_2 } \right)$,\\  $v_2^0  =  - \cos \left( {\pi x_1 } \right)\sin \left( {\pi x_2 } \right)$ and $f_1=f_2=0$.\\\\ Substituting for $v_1^0$ and $v_2^0$ in equations $\left(3\right)$ and $\left(18\right)$ we get
%
\begin{eqnarray}
g_i^0\!=\!\frac{1}
{{2\pi }}\!\int\limits_{\mathbb{R}^2 }\! {\frac{{\left( {x_i  - y_i } \right)\sum\limits_{k = 1}^2 {\frac{{\partial g_k^0 \left( {y,t} \right)}}
{{\partial y_ik}}} }}
{{P_2 \left( {x,y} \right)}}} \prod\limits_{j = 1}^n {dy_j } \! =\! \frac{\pi }
{2}\sin \left( {2\pi x_i } \right)\qquad
\end{eqnarray}
resulting in $\mathcal{U}_i^0\left( v^0\right)=\mathcal{U}_i\left( v \right)\equiv 0$. We obtain $v\left( {x,t} \right)$, $p\left( {x,t} \right)$ from  equations $\left(17\right)$ and $\left(7\right)$:
%
\begin{eqnarray}
v_1  = \sin \left( {\pi x_1 } \right)\cos \left( {\pi x_2 } \right)e^{ - 2\pi ^2 \kappa t}
\end{eqnarray}
%
\begin{eqnarray}
v_2  =  - \cos \left( {\pi x_1 } \right)\sin \left( {\pi x_2 } \right)e^{ - 2\pi ^2 \kappa t}
\end{eqnarray}
and
\begin{eqnarray}
p =  - \frac{{\rho e^{ - 4\pi ^2 \kappa t} }}{4}\left[ {\cos \left( {2\pi x_1 } \right) + \cos \left( {2\pi x_2 } \right)} \right]
\end{eqnarray}
which is the two dimensional Taylor vortex solution\cite{Tay}. The identities used to evaluate the integrals in equation $\left(17\right)$ may be found in Gradshteyn and Ryzhik\cite{gra}.\\\\
\protect{\small{\bfseries{Problem$\left(\rm{ii}\right)$}}} $v_1^0  = \sin \left( {\pi x_1 } \right)\cos \left( {\pi x_2 } \right)$,\\$v_2^0  =  - \cos \left( {\pi x_1 } \right)\sin \left( {\pi x_2 } \right)$ and $f_1  = \mathcal{G}\left( t \right)v_1^0$,\\ $f_2  = \mathcal{G}\left( t \right)v_2^0$. $\mathcal{G}\left( t \right)$ a function of time only and ${e^{ - 2\pi ^2 \kappa t} \int_0^t {\mathcal{G}\left( \tau  \right)e^{2\pi ^2 \kappa \tau } } d\tau }$ is integrable in any given interval.\\\\ We obtain $v\left( {x,t} \right)$, $p\left( {x,t} \right)$ from  equations $\left(15\right)$ and $\left(7\right)$:
%
\begin{eqnarray}
v_1  = \sin \left( {\pi x_1 } \right)\cos \left( {\pi x_2 } \right)\Omega \left( t \right)e^{ - 2\pi ^2 \kappa t} 
\end{eqnarray}
%
\begin{eqnarray}
v_2  =  - \cos \left( {\pi x_1 } \right)\sin \left( {\pi x_2 } \right)\Omega \left( t \right)e^{ - 2\pi ^2 \kappa t} 
\end{eqnarray}
and
%
\begin{eqnarray}
p =  - \frac{{\rho \Omega ^2 \left( t \right)e^{ - 4\pi ^2 \kappa t} }}
{4}\left[ {\cos \left( {2\pi x_1 } \right) + \cos \left( {2\pi x_2 } \right)} \right]\quad
\end{eqnarray}
where 
\begin{eqnarray}
\Omega \left( t \right) = 1 + \int_0^t {\mathcal{G}\left( \tau  \right)e^{2\pi ^2 \kappa \tau } } d\tau 
\end{eqnarray} 
We note that the solution satisfies the necessary condition given by equation $\left(18\right)$
%
\begin{eqnarray}
g_i & =& \frac{1}
{{2\pi }}\int\limits_{\mathbb{R}^2 } {\frac{{\left( {x_i  - y_i } \right)\sum\limits_{k = 1}^2 {\frac{{\partial g_k \left( {y ,t} \right)}}
{{\partial y_k }}} }}
{{P_2 \left( {y_j ,x_j } \right)}}} \prod\limits_{j = 1}^n {dy_j } 
\nonumber\\
&  = &  \frac{\pi }{2}\sin \left( {2\pi x_i } \right)\Omega ^2 \left( t \right)e^{ - 4\pi ^2 \kappa t} ,\qquad i=1,2
\end{eqnarray}
resulting in the umbilical force $\mathcal{U}_i\left( v \right)\equiv 0$.
\begin{center}
\protect{{\bfseries{Exact solutions in $\mathbb{R}^3$\\ $\mathbb{R}^3=\left\{ { - \infty  < x_i  < \infty ; \,\, i = 1,2,3} \right\}$}}}
\end{center}
\protect{\small{\bfseries{Problem$\left(\rm{i}\right)$}}} $v_1^0  = a\sin \pi x_3  - c\cos \pi x_2$,\\ $v_2^0\!  =\! b\sin \pi x_1 - a\cos \pi x_3$, $v_3^0 \! = \!c\sin \pi x_2  - b\cos \pi x_1$
 and $f_1=f_2=f_3=0$. $a$, $b$ and $c$ are real constants.\\\\ Substituting for $v_1^0$, $v_2^0$ and $v_3^0$ in equations $\left(3\right)$ and $\left(18\right)$ we get
%
\begin{eqnarray}
g_1^0  &=& \frac{1}
{{4\pi }}\int\limits_{\mathbb{R}^3 } {\frac{{\left( {x_1  - y_1 } \right)\sum\limits_{k = 1}^3 {\frac{{\partial g_k^0 \left( {y ,t} \right)}}
{{\partial y_k }}} }}
{{\left\{ {\mathcal{P}_3\left( {x,y} \right)} \right\}^{\frac{3}
{2}} }}} \prod\limits_{j = 1}^3 {dy_j } 
\nonumber\\
&  = & 
\pi\! \left\{ {bc\sin \left(\pi x_1\right) \sin \left(\pi x_2\right)\!  - \!ab\cos \left(\pi x_1\right) \cos\left( \pi x_3\right) } \!\right\}\qquad
\end{eqnarray}
%
%
\begin{eqnarray}
g_2^0  &  = & \frac{1}
{{4\pi }}\int\limits_{\mathbb{R}^3 } {\frac{{\left( {x_2  - y_2 } \right)\sum\limits_{k = 1}^3 {\frac{{\partial g_k^0 \left( {y ,t} \right)}}
{{\partial y_k }}} }}
{{\left\{ {\mathcal{P}_3\left( {x,y} \right)} \right\}^{\frac{3}
{2}} }}} \prod\limits_{j = 1}^3 {dy_j } 
\nonumber\\
&  = &\pi\! \left\{ {ac\sin \left(\pi x_2\right) \sin \left(\pi x_3\right) \! -\! bc\cos \left(\pi x_1\right) \cos \left(\pi x_2\right) } \!\right\}\qquad
\end{eqnarray}
%
\begin{eqnarray}
g_3^0 &  = &\frac{1}
{{4\pi }}\int\limits_{\mathbb{R}^3 } {\frac{{\left( {x_3  - y_3 } \right)\sum\limits_{k = 1}^3 {\frac{{\partial g_k^0 \left( {y,t} \right)}}
{{\partial y_k}}} }}
{{\left\{ {\mathcal{P}_3\left( {x,y } \right)} \right\}^{\frac{3}
{2}} }}} \prod\limits_{j = 1}^3 {dy_j } 
\nonumber\\
&  = & \pi \!\left\{ {ab\sin\left( \pi x_1\right) \sin \left(\pi x_3\right)\!  -\! ac\cos\left( \pi x_2\right) \cos \left(\pi x_3\right) } \!\right\}\qquad
\end{eqnarray}
Hence, the umbilical force $\mathcal{U}_i^0\left( v^0\right)=\mathcal{U}_i\left( v \right)\equiv 0$. We obtain $v\left( {x,t} \right)$, $p\left( {x,t} \right)$ from  equations $\left(17\right)$ and $\left(7\right)$:
%
\begin{eqnarray}
v_1 = \left\{ {a\sin \left( {\pi x_3 } \right) - c\cos \left( {\pi x_2 } \right)} \right\}e^{ - \pi ^2 \kappa t} 
\end{eqnarray}
%
\begin{eqnarray}
v_2  = \left\{b\sin \left( {\pi x_1 } \right)- {a\cos \left( {\pi x_3 } \right)} \right\}e^{ - \pi ^2 \kappa t} 
\end{eqnarray}
%
\begin{eqnarray}
v_3  = \left\{ {c\sin \left( {\pi x_2 } \right) - b\cos \left( {\pi x_1 } \right)} \right\}e^{ - \pi ^2 \kappa t} 
\end{eqnarray}
and
\begin{eqnarray}
p &= & - \rho e^{ - 2\pi ^2 \kappa t} \left[ {bc\cos \left( {\pi x_1 } \right)\sin \left( {\pi x_2 } \right) + } \right.\nonumber\\
&+&\left. {ab\cos \left( {\pi x_3 } \right)\sin \left( {\pi x_1 } \right) + ac\cos \left( {\pi x_2 } \right)\sin \left( {\pi x_3 } \right)} \right]\qquad
\end{eqnarray}
which is the non-stationary solution of the Navier-Stokes equation for the Arnold-Beltrami-Childress (ABC) flow previously studied by Dombre et al.\cite{Dom} for stationary solution of Euler's equation.\\\\
\protect{\small{\bfseries{Problem$\left(\rm{ii}\right)$}}} $v_1^0  = a\sin \pi x_3  - c\cos \pi x_2$,\\ $v_2^0\!  =\!  b\sin \pi x_1- a\cos \pi x_3$, $v_3^0 \! = \!c\sin \pi x_2  - b\cos \pi x_1$ and $f_1  = \mathcal{G}\left( t \right)v_1^0$, $f_2  = \mathcal{G}\left( t \right)v_2^0$, $f_3  = \mathcal{G}\left( t \right)v_3^0$. $\mathcal{G}\left( t \right)$ a function of time only.\\\\ We obtain $v\left( {x,t} \right)$, $p\left( {x,t} \right)$ from  equations $\left(15\right)$ and $\left(7\right)$:
%
\begin{eqnarray}
v_1  = \left\{ {a\sin \left( {\pi x_3 } \right) - c\cos \left( {\pi x_2 } \right)} \right\}\Omega \left( t \right)e^{ - \pi ^2 \kappa t} 
\end{eqnarray}
%
\begin{eqnarray}
v_2  = \left\{ {b\sin \left( {\pi x_1 } \right) - a\cos \left( {\pi x_3 } \right)} \right\}\Omega \left( t \right)e^{ - \pi ^2 \kappa t} 
\end{eqnarray}
%
\begin{eqnarray}
v_3  = \left\{ {c\sin \left( {\pi x_2 } \right) - b\cos \left( {\pi x_1 } \right)} \right\}\Omega \left( t \right)e^{ - \pi ^2 \kappa t} 
\end{eqnarray}
and
%
\begin{eqnarray}
p &= & - \rho \Omega ^2 \left( t \right)e^{ - 2\pi ^2 \kappa t} \left[ {bc\cos \left( {\pi x_1 } \right)\sin \left( {\pi x_2 } \right) + } \right.\nonumber\\
&+&\left. {ab\cos \left( {\pi x_3 } \right)\sin \left( {\pi x_1 } \right) + ac\cos \left( {\pi x_2 } \right)\sin \left( {\pi x_3 } \right)} \right]\qquad
\end{eqnarray}
where 
\begin{eqnarray}
\Omega \left( t \right) = 1 + \int_0^t {\mathcal{G}\left( \tau  \right)e^{\pi ^2 \kappa \tau } } d\tau 
\end{eqnarray} 
We note that the solution satisfies the necessary condition given by equation $\left(18\right)$
%
%
\begin{eqnarray}
g_1  &=& \frac{1}
{{4\pi }}\int\limits_{\mathbb{R}^3 } {\frac{{\left( {x_1  - y_1 } \right)\sum\limits_{k = 1}^3 {\frac{{\partial g_k \left( {y ,t} \right)}}
{{\partial y_k }}} }}
{{\left\{ {\mathcal{P}_3\left( {x ,y} \right)} \right\}^{\frac{3}
{2}} }}} \prod\limits_{j = 1}^3 {dy_j } 
\nonumber\\
&  = &\! 
\pi\! \left\{ {bc\sin \pi x_1 \sin \pi x_2\!  - \!ab\cos \pi x_1 \cos \pi x_3 } \right\}\!\Omega^2\! \left( t \right)\!e^{ - 2\pi ^2\! \kappa t} \nonumber\\
\end{eqnarray}
%
%
\begin{eqnarray}
g_2  &  = & \frac{1}
{{4\pi }}\int\limits_{\mathbb{R}^3 } {\frac{{\left( {x_2  - y_2 } \right)\sum\limits_{k= 1}^3 {\frac{{\partial g_k \left( {y,t} \right)}}
{{\partial y_k }}} }}
{{\left\{ {\mathcal{P}_3\left( {x ,y} \right)} \right\}^{\frac{3}
{2}} }}} \prod\limits_{j = 1}^3 {dy_j } 
\nonumber\\
&  = &\!\pi\! \left\{ {ac\sin \pi x_2 \sin \pi x_3\!  -\! bc\cos \pi x_1 \cos \pi x_2 } \right\}\!\Omega^2\! \left( t \right)\!e^{ - 2\pi ^2\! \kappa t} \nonumber\\
\end{eqnarray}
%
\begin{eqnarray}
g_3 &  = &\frac{1}
{{4\pi }}\int\limits_{\mathbb{R}^3 } {\frac{{\left( {x_3  - y_3 } \right)\sum\limits_{k = 1}^3 {\frac{{\partial g_k \left( {y,t} \right)}}
{{\partial y_k}}} }}
{{\left\{ {\mathcal{P}_3\left( {x ,y } \right)} \right\}^{\frac{3}
{2}} }}} \prod\limits_{j = 1}^3 {dy_j } 
\nonumber\\
&  = &\! \pi\! \left\{ {ab\sin \pi x_1 \sin \pi x_3\!  -\! ac\cos \pi x_2 \cos \pi x_3 } \right\}\!\Omega^2\! \left( t \right)\!e^{ - 2\pi ^2\! \kappa t} \nonumber\\ 
\end{eqnarray}
resulting in the umbilical force $\mathcal{U}_i\left( v \right)\equiv 0$.\\\\ 
\protect{\small{\bfseries{Problem$\left(\rm{iii}\right)$}}}  $v_1^0  = a\sin \pi x_3  - c\cos \pi x_2$,\\ $v_2^0\!  =\!  b\sin \pi x_1- a\cos \pi x_3$, $v_3^0 \! = \!c\sin \pi x_2  - b\cos \pi x_1$ and 
$f_1  = f_I e^{ - \lambda t}$, $f_I$ and $\lambda$ are constants, $f_2  = f_3  = 0$.\\\\ The umbilical force for this case is zero. We obtain $v\left( {x,t} \right)$, $p\left( {x,t} \right)$ from  equations $\left(15\right)$ and $\left(7\right)$:
\begin{eqnarray}
v_1  &=& \left\{ {a\sin \left( {\pi x_3 } \right) - c\cos \left( {\pi x_2 } \right)} \right\}e^{ - \pi ^2 \kappa t}  + \nonumber\\
&+&\left\{ {\begin{array}{*{20}c}
   {f_I t}, & {\lambda  = 0}  \\
   {\frac{{f_I }}
{\lambda }\left( {1 - e^{ - \lambda t} } \right)}, & {\lambda  > 0}  \\
 \end{array} } \right.
\end{eqnarray}
$v_2$, $v_3$ and $p$ are given by equations (32), (33) and (34) respectively.
\section{Concluding Remarks} 
Exact analytical solutions in $\mathbb{R}^2$ and $\mathbb{R}^3$ have been developed for the case where the nonlinear umbilical force vanishes by construction. It is shown that, if the initial conditions and the components of the externally applied force are chosen such that the umbilical force is zero and incompressibility of the fluid is enforced at all times, then the smooth solution of the diffusion equation can also be a solution of the Navier-Stokes equation.\\\\ The solutions presented here satisfy $\left(1\right)$, $\left(2\right)$, $\left(4\right)$, $\left(5\right)$ and $p,v \in C^\infty  \left( {\mathbb{R}^n  \times \left[ {0,\infty )} \right.} \right)$. For all solutions presented here, the vorticity may be obtained from the curl of the velocity.  
\begin{acknowledgments}
The author would like to thank Tarek Habashy and Jeff Spath for many valuable discussions. We acknowledge, with thanks, the conversations we have had with Shalini Krishnamurthy who gave her time to read and critique this manuscript in many of its avatars. Helpful comments from Wentao Zhou, Peter Tilke, Greg Grove and Tom Bratton are also acknowledged.
\end{acknowledgments}

\bibliography{apssamp}
\end{document}